\documentclass{aa}  

\usepackage{graphicx}
\usepackage{txfonts}
\begin{document}

   \title{No evidence for absence of solar dynamo synchronization}


   \author{F. Stefani \inst{1}, J. Beer \inst{2}, T. Weier
          \inst{1}
          }

   \institute{Helmholtz-Zentrum Dresden - Rossendorf, Bautzner Landstr. 400, 01328 Dresden, Germany\\
              \email{F.Stefani@hzdr.de, T.Weier@hzdr.de}
         \and
             Eawag, \"Uberlandstrasse 133, 8600 D\"ubendorf, Switzerland\\
             \email{Juerg.Beer@eawag.ch}
             }

   \date{Received September 15, 1996; accepted March 16, 1997}

 
  \abstract
   {The old question of whether the solar dynamo is synchronized by the
   tidal forces of the orbiting planets has recently received renewed interest, both 
   from the viewpoint of historical data analysis and in terms of theoretical and numerical modelling.}
   {We aim to contribute to the solution of this longstanding puzzle by analyzing cosmogenic radionuclide data from the last millennium.}
   {We reconsider a recent time-series of $^{14}$C-inferred sunspot data and compare the resulting cycle minima and maxima with the corresponding conventional series down to 1610 A.D., enhanced by Schove's data before that time.}
   {We find that, despite recent claims to the contrary, the 
   $^{14}$C-inferred sunspot data are well compatible with a synchronized solar dynamo, exhibiting
   a relatively phase-stable period of 11.07 years, which points to a synchronizing role of the spring tides of the Venus-Earth-Jupiter system.}
   {}

   \keywords{Sun: activity --- Sun: dynamo
               }

   \maketitle
%

\section{Introduction}

   The question of whether the solar dynamo might be ``clocked'' by the motion of the planets traces back to
   early speculations by \cite{Wolf1859}, and has popped up sporadically ever since (\cite{delarue1872},
   \cite{Bolling1952}, \cite{Jose1965}, \cite{Takahashi1968}, \cite{Wood1972}, \cite{Dejager2005}).
   Recently, new impetus was given to the issue by 
   the exemplification of \cite{Hung2007}, \cite{Scafetta2012}, \cite{Wilson2013}, and \cite{Okhlopkov2016} that the 11.07-yr spring-tide period of the tidally dominant planets Venus, Earth and Jupiter appears to be in a phase-stable relation with the solar cycle. This
   finding turned out to be in amazing agreement with
   the older results of Schove's  ambitious ``spectrum of time'' project 
   (Schove \citeyear{Schove1983}) to determine the solar cycle maxima and minima for the last two and a half millennia mainly from historical {\it aurora borealis} sightings and naked-eye sunspot observations. Furthermore, the identified 11.07-yr periodicity is also, within the error margin,
   well compatible with the phase-stable 11.04-yr cycle as inferred by \cite{Vos2004} utilizing two different algae data-sets from the early Holocene.
 
   The key problem with those observed correlations is how they could be substantiated by any kind of a viable causation. While the tidal forces of the planets can easily be ridiculed by the minuscule tidal height of the order of 1\,mm, several physical mechanisms have been invoked that could possibly lead to noticeable effects, among them the extreme sensitivity of the storage capacity for magnetic fields in the sub-adiabatic tachocline (Abreu et al. \citeyear{Abreu2012}, Charbonneau \citeyear{Charbonneau2022}), or the susceptibility of intrinsic helicity oscillations of waves or instabilities (in particular, the Tayler instability) to tidal forces (Weber et al. \citeyear{Weber2013}, Weber et al. \citeyear{Weber2015}, Stefani et al. \citeyear{Stefani2016}, Stefani et al. \citeyear{Stefani2019}, Stefani et al. \citeyear{Stefani2020a}, Stefani et al. \citeyear{Stefani2021}). 

   Going beyond such mainly qualitative arguments, the recent work of \cite{Horstmann2023} has shown that even weak tidal forces such as of Jupiter might excite (magneto)-Rossby waves with typical velocity amplitudes of up to m/s. A concurrent 2-dimensional simulation by \cite{Klevs2023} affirmed that tidally triggered oscillations of the tachoclinic $\alpha$ effect of the order of dm/s would be sufficient to synchronize an otherwise conventional $\alpha-\Omega$ dynamo model. Together with the older argument of \cite{Opik1972} that the ``ridiculous'' 1\,mm tidal height corresponds energetically to a velocity scale of 1 m/s, those recent results exemplify that tidal forces may entail a serious potential for solar dynamo synchronization.

   Still, a hard-to-solve problem of that kind would not even appear if the solar dynamo was not phase-stable in the first place. Two recent papers (Nataf \citeyear{Nataf2022}, Weisshaar et al. \citeyear{Weisshaar2023}) have seriously put into question the empirical evidence for phase-stability. Ignoring the strong argument in favour of phase stability coming from the algae-date in the early Holocene (Vos \citeyear{Vos2004}), both papers focus exclusively on the series of solar cycle extrema (i.e., minima or maxima) during the last millennium. \citet{Nataf2022} dismissed all the meticulous efforts of \cite{Schove1983} by claiming his cycle reconstruction to be construed by simple rules.  Strictly presupposing  that the solar cycle {\it is not} clocked, he argued that Schove's ``nine-per-century'' rule would lead to a constrained, and therefore wrong, series of extrema. What was not considered by \cite{Nataf2022}, though, was the possibility that the solar cycle {\it might indeed be} clocked by a 11-07-yr trigger, in which case Schove's {\it auxiliary} ``nine-per-century'' rule would do absolutely no harm to an otherwise correctly inferred series of extrema. In this respect it is interesting to note that Schove's data actually point to an average Schwabe cycle of 11.07 years rather than 11.11 years which would result from  a naive application of  the ``nine-per century'' rule (see Fig. 1 of \cite{Stefani2020a}). For further critical remarks on \citet{Nataf2022}, see the recent comment by \cite{Scafetta2023}.

   \begin{figure*}
    \sidecaption
   \includegraphics[width=0.53\textwidth]{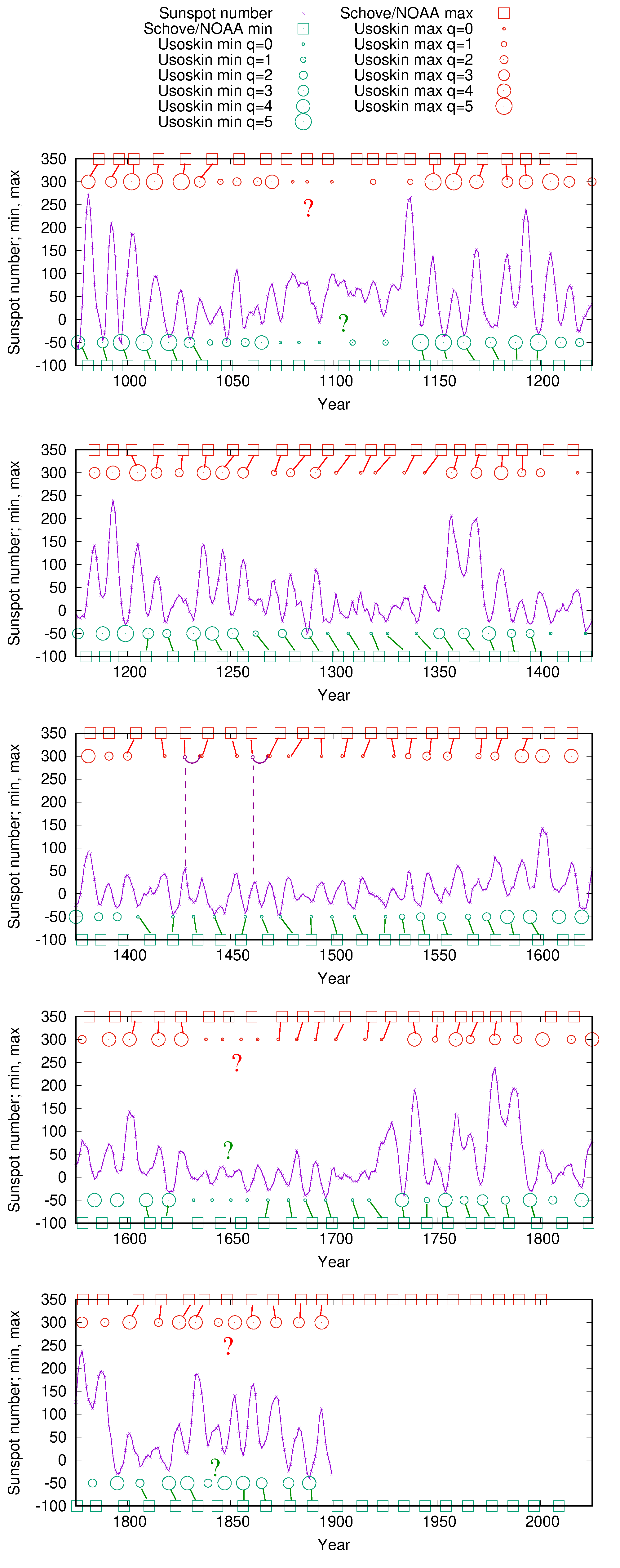}
   \caption{Series of annual (pseudo) sunspot numbers (violet lines) for the time interval between 975 and 1895, 
   and inferred solar cycle maxima (red circles) and minima (green circles) according to \cite{Usoskin2021}, together with a merger of the maxima (red squares) and minima (green squares)
   of \cite{Schove1983} and NOAA (https://www.ngdc.noaa.gov/stp/solar/solardataservices.html). Each individual panel shows a period of
   250 years, with overlapping intervals of 25 years at 
   either margin. 
   Red and green thin lines
   indicate putative correspondences
   between the respective extrema (they are usually restricted  to the central 200-yr interval of the corresponding panel, except in the first panel). The size of the red and green circles 
   mirrors the quality flag $q$ according to \cite{Usoskin2021} (smallest circles denote very poor quality, $q=0$, largest circles denote highest quality, $q=5$). 
   Red and green question marks indicate intervals with 
   unclear correspondences, typically at times characterized by low 
   quality flags. The two dashed violet lines around 1425 and 1461 
   indicate corrections of obvious
   typing errors for the maxima in Table 1 of \cite{Usoskin2021}.
   After these two trivial corrections, uninterrupted 
   one-to-one matches between the extrema of both series become visible
   between 1140 and 1620.
   }
              \label{FigGam1}%
    \end{figure*}

A second paper that claims to have finally debunked the clocking scenario was recently published by \cite{Weisshaar2023}. Based on $^{14}$C data of \cite{Brehm2021} for the last millennium, it uses the series of cycle extrema inferred by \cite{Usoskin2021} to show that this series points - with a high statistical significance - to a random walk process instead of a clocked process.

    In the present paper we will reanalyze this series of cycle minima and maxima and compare it with another series of extrema, for which we use a combination  of the standard Schwabe cycles for the later time interval starting at 1610  with Schove's data (Schove \citeyear{Schove1983}) for earlier times. We will show that - for the most part of the interval - the extrema of both series can uniquely be matched one-to-another with three exceptions. The latest of those shows up around 1840 where Usoskin's data  exhibit two shallow minima at a place where the telescopic data show only one. Given the shallowness of these two minima, and the relatively low quality flag of the first one, we find it legitimate to replace this pair by only one minimum, and to cancel the corresponding maximum between them. A second ambiguity is found amidst the Maunder minimum around 1650 where all quality flags of Usoskin's data are relatively low. Here again we cancel one shallow minimum. The most problematic part appears in the interval between 1040 and 1140 where Usoskin's quality flags are generally quite low. Specifically, we consider three different ways of correcting the data which we all consider at least as plausible as the original selection of 
 \cite{Usoskin2021}.

   Then, we will analyze the considered times series with view on their phase stability. At first we show the respective Observed-minus-Calculated (O-C) diagrams of the residuals of the instants of the minima from a theoretical linear trend with an alleged 11.07-yr period. While the Schove/NOAA data are concentrated around a horizontal line with only slight ($\pm$ 4 years) upward or downward deviations, Usoskin's data show larger deviations exactly within the three problematic intervals discussed above. We show how these deviations are consecutively reduced by our corrections. Thereby we arrive at a one-to-one matching of the $^{14}$C extrema with those of the Schove/NOAA series for the entire 9 century long interval. 

   In the last step we compute - for the different time series - Dicke's ratio between the standard deviation of the residuals and the standard deviation of the differences between neighbouring residuals. A closely related measure, defined  by \citet{Gough1981}, was used by \cite{Weisshaar2023} and \cite{Biswas2023} to argue in favour of a random walk process. We show here that already the two highly plausible corrections around 1840 and 1650 lead to a dramatic move of Dicke's ratio towards the corresponding theoretical curve for a clocked process. 
   Finally we show that a good deal of the remaining deviations from a strictly clocked curve turns out to be due to the presence of a well-expressed Suess-de Vries cycle in the data.

   The paper closes with some conclusions.


\section{Data sets and possible corrections}

In the following, we will discuss two data sets. The first one is the annular series of (pseudo) sunspot numbers as recently inferred by \cite{Usoskin2021} from
the $^{14}$C production rate data of \cite{Brehm2021}. To start with, we simply
adopt the cycle minima and maxima as derived by  \cite{Usoskin2021}.

The second sequence of minima and maxima consists of the standard data from NOAA with the starting year 1610
(https://www.ngdc.noaa.gov/stp/solar/solardataservices.html), 
merged with the earlier data as published by \cite{Schove1983} and partly
corrected in \cite{Schove1984}.

In Fig. 1, these data are presented in five 200-yr intervals, with 25-yr overlaps at each beginning and end.
The size of the symbols of Usoskin's minima (green open circles) and maxima
(red open circles) is scaled by the so-called ``quality flag'' between 
$q=0$ (``cycle cannot be reliable identified'') till $q=5$
(``clear cycle in both shape and amplitude''). The violet line depicts
the (pseudo) sunspot number according to \cite{Usoskin2021}, where in most cases the attribution to the minima and maxima is rather clear. 
Two evident exceptions occur for the maxima at 1435 and 1468, which
are obviously
due to typing errors in Table 1 of \cite{Usoskin2021}, and which we correct, according to the $^{14}$C-curve (violet dashed lines in the middle panel of Fig. 1) to the years 1425 and 1461, respectively.

The red and green open squares at the upper and lower abscissa depict the 
Schove/NOAA maxima and minima, respectively.
After the two trivial corrections mentioned above, we obtain a sequence of one-to-one matches between the 44 minima and maxima of Usoskin and Schove/NOAA that stretches uninterruptedly over the time interval between 1140 and 1620.

Evidently, there are three distinctive segments where this one-to-one correspondence fails or at least becomes problematic. The first one concerns the long interval between 1040 and 1140, the second one is situated around 1650, the third one around 1840.

   
%

 \begin{figure}
   \sidecaption
   \includegraphics[width=0.49\textwidth]{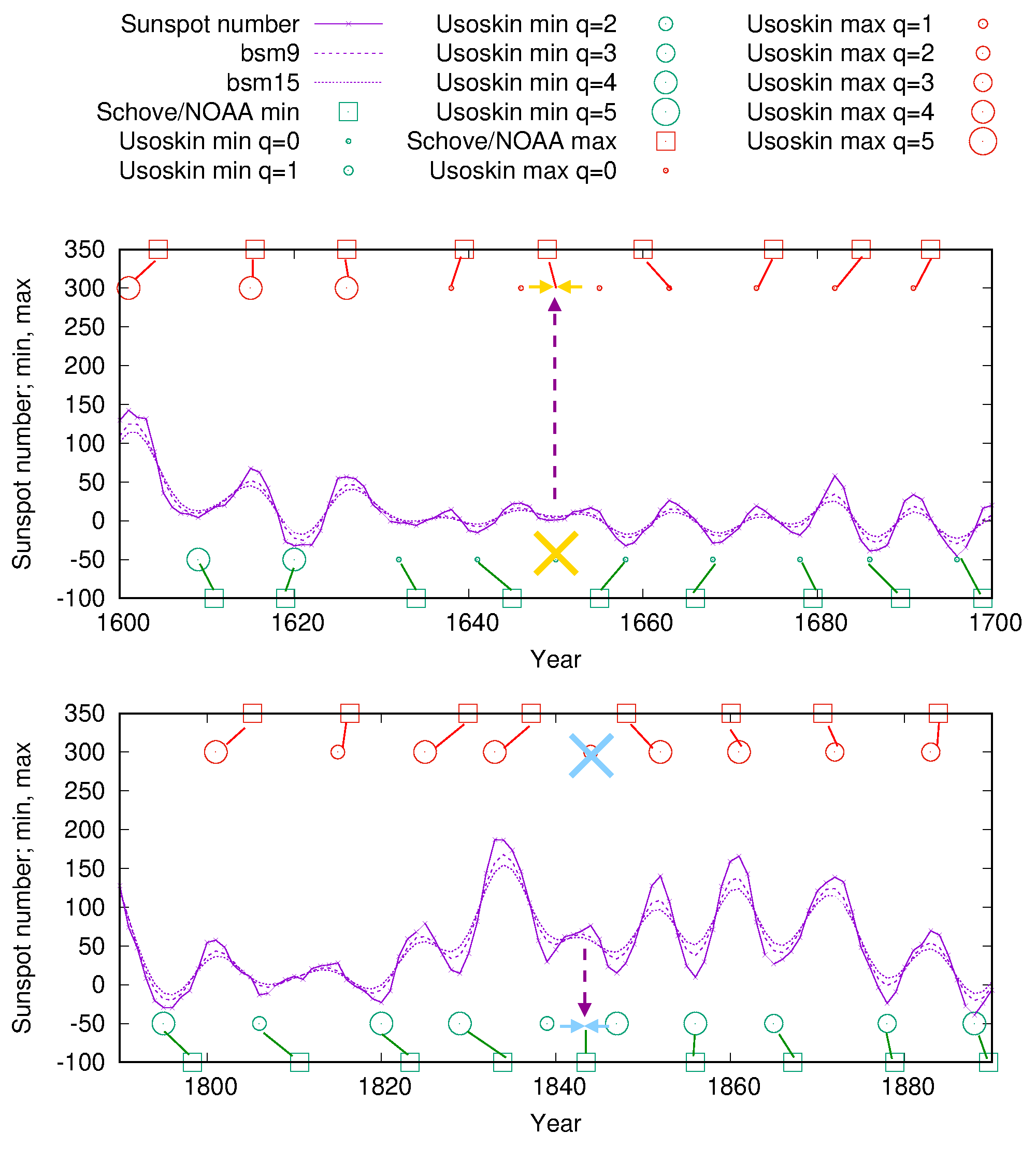}
   \caption{Plausible corrections of minima/maxima in two late time intervals. Symbols and lines as in Fig. 1., except that two binomially filtered curves  bsm 9 (including 9 coefficients, dashed violet) and bsm 15 (including 15 coefficients, dotted violet) are added to the original $^{14}$C data (violet full line). Lower panel: Interval from 1790 till 1890. The filtered curves insinuate that the two minima at 1839 (with a low quality flag of $q=2$) and 1846 might indeed be only one minimum at 1843 which would also better fit to the observational data. The merging of the minima is indicated by two horizontal light-blue arrows, the cancellation of the maxima between them by a corresponding light blue cross. Upper panel: Interval between 1600 and 1700. Here, amidst the Maunder minimum, the quality flags of the minima and maxima are typically low, making their unambiguous identification quite hard. A most plausible cancellation of a flat minimum is indicated by a yellow cross.}
              \label{FigGam2}%
    \end{figure}

   \begin{figure}
   \sidecaption
   \includegraphics[width=0.49\textwidth]{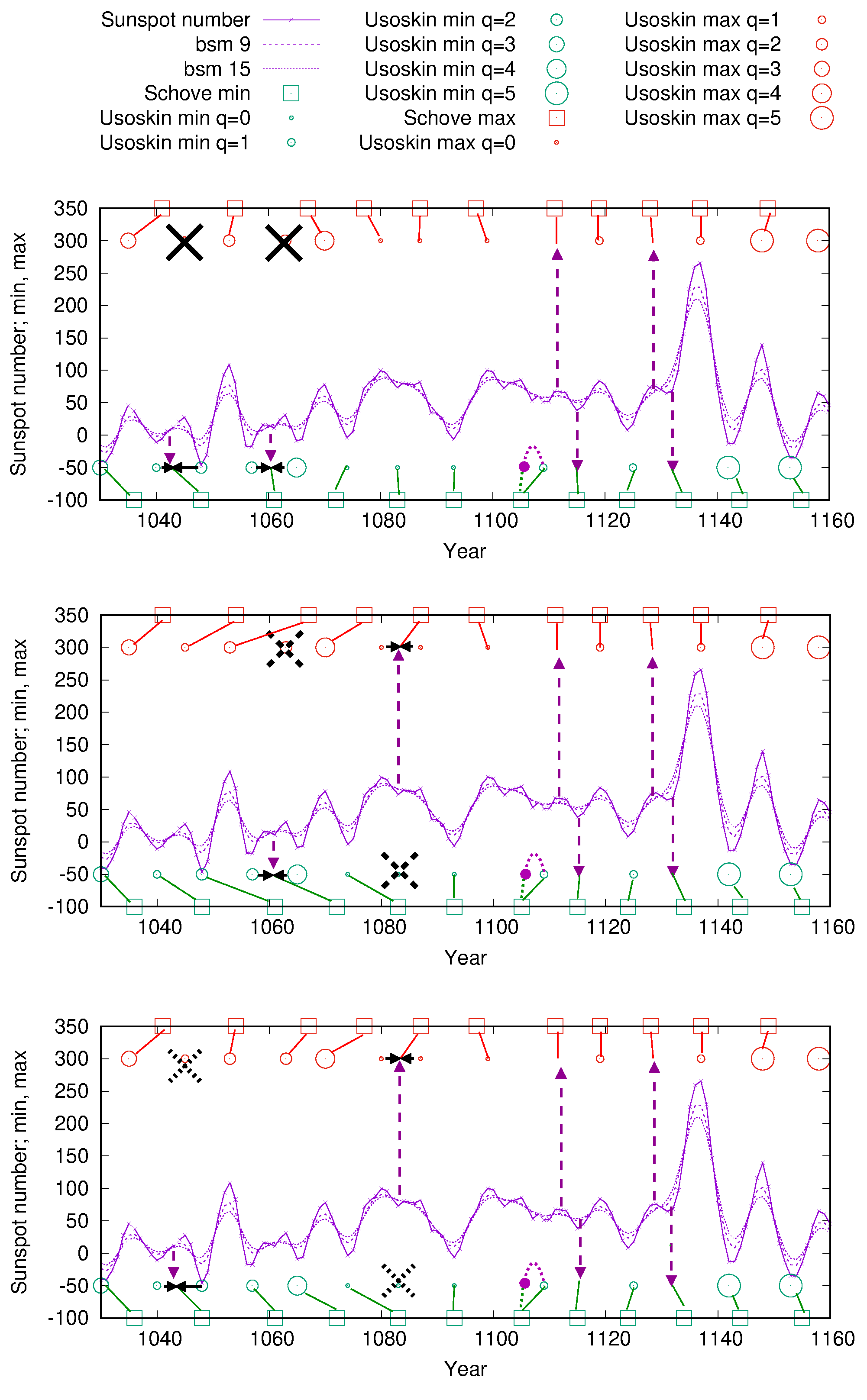}
   \caption{As Fig. 2, but for the early interval between 1030 and 1160.
   The vertical violet lines indicate the insertion of two minima/maxima pairs in the later segment of that interval. The black horizontal arrows and crosses indicate contractions and 
   cancellations of various minima or maxima in the early segment of the interval, of which we show 
   three different permutations indicated by full, dashed and dotted types of black crosses.}
              \label{FigGam3}%
    \end{figure}

In Fig. 2 we consider two particular segments in more detail. Let us start with the latest part, around 1840, which is shown in the lower panel of Fig. 2. Evidently, the NOAA data comprise only one minimum at 1843.5, whereas Usoskin's data show two minima here, at 1839 and 1847, the former of which having a relatively low quality flag of $q=2$. In order to shed more light on this issue we add to the original violet curve of sunspot data two further curves representing two different binomial smoothings (bsm 9 and bsm 15) which tend to smear out the two shallow minima into a single one centered at 1843, which indeed corresponds to the NOAA value. Given the high validity of the observationally constrained minimum in the middle of the 19th century, we consider such contraction of two minima into a single one (indicated by the two light blue arrows) and the corresponding cancellation (light blue cross) of one maximum as highly plausible.

Next we turn to the situation around 1645 (upper panel in Fig. 2). Here, amidst the Maunder minimum, we should have less trust into the NOAA data (as for the problem of interpretation of naked-eye sunspot observations during this time, see \cite{Carrasco2020}). Hence, it is not excluded that the additional minimum/maximum pair of Usoskin's data is indeed
a real one. This possibility will be discussed below. Still, it is also plausible that Usoskin's data show one minimum/maximum pair 
too much. From the most evident variants to contract either the two (flat) 
maxima at 1638 and 1646 or those at 1646 and 1655, we show only the latter one, indicated with yellow 
arrows, together with the cancelled minimum at 1650 (yellow cross).

\begin{figure}
     \centering
   \includegraphics[width=0.49\textwidth]{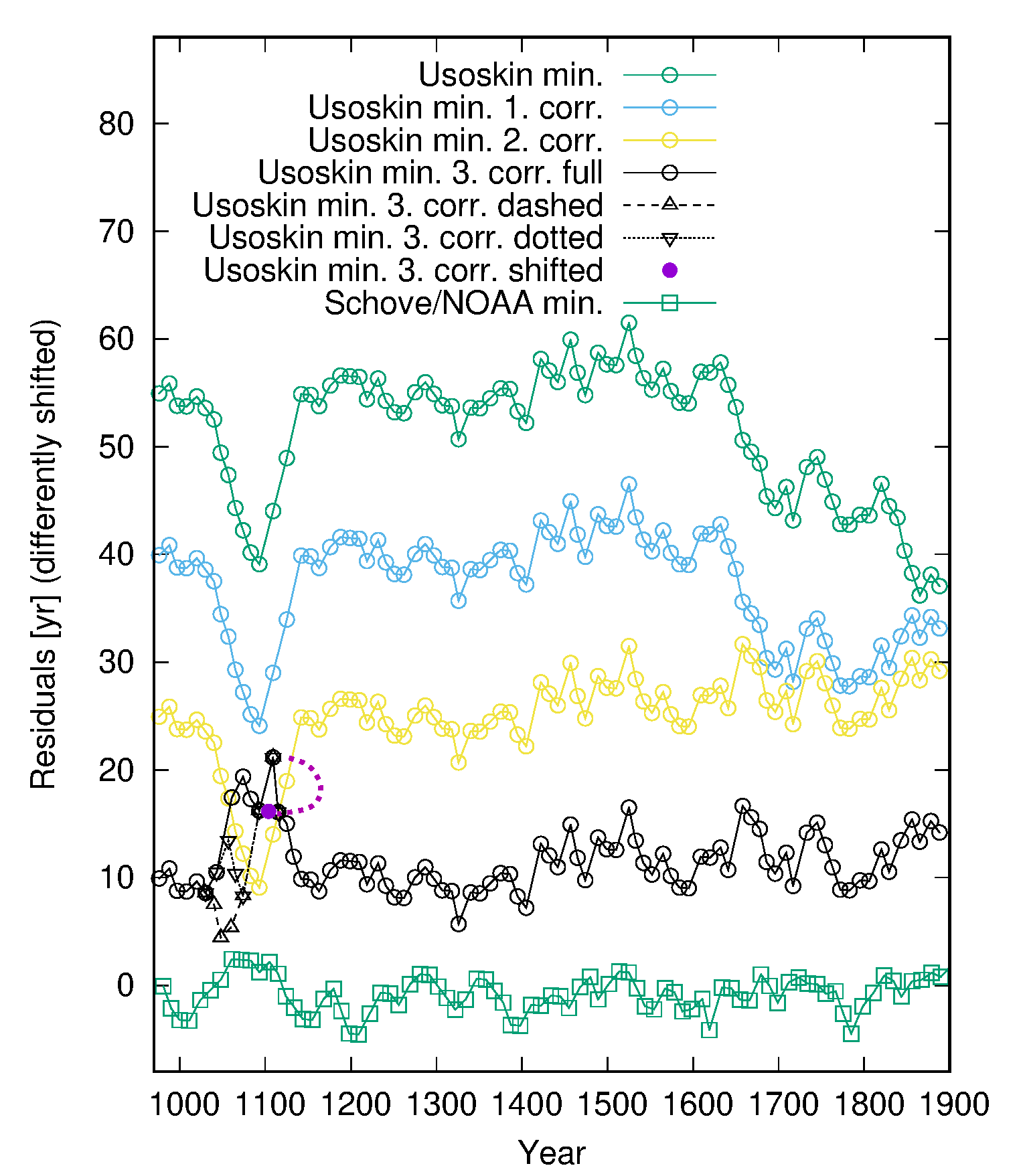}
   \caption{O-C plot of various data sets of cycle minima. For the sake of better visibility, the residuals (which all refer to a linear trend with 11.07-yr period), are differently shifted on the ordinate axis.  The light-blue, yellow, and black curves correspond to the different corrections in the various panels of Figs. 2 and 3. The single violet full circle corresponds to a possible shift of the 1109 year minimum as proposed in Fig. 3.}
              \label{FigGam4}%
    \end{figure}

Finally, we treat, in Fig. 3, the long period between 
1040 and 1140. This interval, which strongly overlaps with the Oort minimum, is characterized by a large number of low quality flags.
While the total number of minimum/maximum pairs in this segment is the same for the 
data of Usoskin and Schove, we immediately notice the presence of two
extremely long neighboring cycles between the maxima at 1099 and 1119 (20 years) 
and 1119 and 1137 (18 years). In either of those intervals we observe the existence 
of one (or two) local minimum/maximum pair(s). Without overemphasizing the validity of Schove's data
(at least the corresponding maxima were not labelled as uncertain by him, in contrast to many others in \cite{Schove1983}),
we see at least that an insertion of the two additional maximum/minimum pairs 
(indicated by the violet arrows) leads to a one-to-one match of the extreme of both 
data sets. Yet, the insertion of the first maximum leads to an 
even greater uncertainty for the
minimum just before it, which might well shift from its place at 1109 (according to Usoskin) to some position before. This variant is indicated by the bent dotted violet curve ending 
in the alternative minimum at 1104 (violet full circle).

Even more uncertain than in this late segment of the 1040-1140 interval is the situation in the early segment which contains quite a number of low-q extrema, leading to a significant number of possible permutations of 
minimum/maximum pairs, in addition to those chosen by \cite{Usoskin2021}. 
In order to ``make good'' for the two {\it insertions} in the later segment, we 
opted here for a compensating {\it cancellation} of two minima/maxima pairs in the early segment, a choice which is, admittedly, strongly debatable. The different panels in Fig. 3 illustrate three most plausible permutations of minimum/maximum cancellations in this early segment. The corresponding 
contractions or cancellations are indicated by black arrows and three types of black crosses 
(full, dashed, dotted). While in all three cases we obtain a one-to-one match of the resulting
minimum/maximum pairs with those of Schove, the first and third variant (with the full and dotted crosses) lead to
the most plausible correspondences.

    \section{O-C plots, Dicke's ratio, and the influence of the Suess-de Vries cycle}

      \begin{figure}
   \centering
   \includegraphics[width=0.49\textwidth]{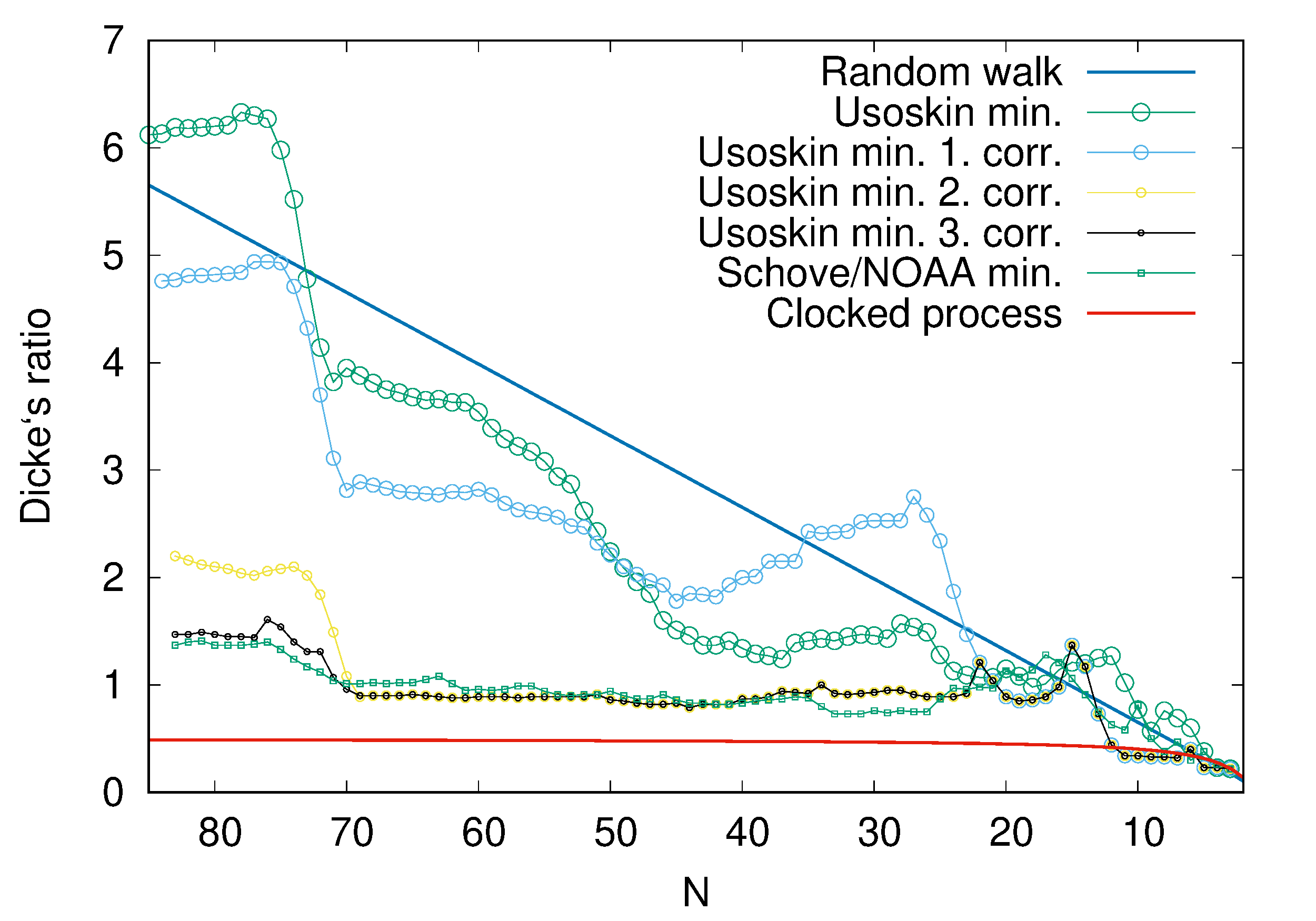}
   \caption{Dicke's ratio for the residuals shown in Fig. 4, with corresponding colors.}
              \label{FigGam5}%
    \end{figure}
    
    A first qualitative hint for phase-stability, or its absence, can be gained from the so-called ``Observed Minus Calculated'' (O-C) diagram, which shows the differences (or residuals) of a given data set from a linear trend for which we use here an alleged
    11.07-yr period of the Schwabe cycle. The lowermost green curve (with open squares) 
    in Fig. 4 shows 
    the residuals for the combined Schove/NOAA minima. Evidently, they are wiggling around a rather horizontal line by typically not more than $\pm 4$ years, which - if confirmed - would
    strongly speak in favour of a noise-perturbed, but nevertheless clocked process. 
    The corresponding residuals for Usoskin's original minima are shown as the uppermost green line with open circles
    (which is - for better visibility - vertically shifted). Between 1140 and 1640 it exhibits already a long horizontal segment, pointing again to phase stability in this interval. Not surprisingly, however, it shows two steep downward-directed phase-jumps around 1650 and 1840, where the two additional minima of \cite{Usoskin2021} are intervening, as discussed earlier. 
    Another remarkable feature is the downward-pointing ``nose'' centered around 1090 which results 
    from the two additional minima in the early segment, combined with two missing minima in the later segment 
    of the 1040-1140 interval.

    The light blue curve represents the residuals (again vertically shifted) after having contracted the two minima at 1839 and 1846 into one at 1843. As a consequence, the downward-directed phase jump disappears here. The next correction, i.e. the cancellation of the minimum at 1650, leads to the yellow curve which is already dominated by a long horizontal segment between 1140 and 1890.

    We now turn to the most problematic ``nose'' between 1040 and 1140. All three black curves in Fig. 4 rely on the insertion of two additional minima at 1115 and 1133 in the late segment, as was specified by the dashed violet arrows in Fig. 3. They differ, however, by the specific combination of the two canceled minima in the early segment, as shown in the three  panels of Fig. 3. 
    Full, dashed, and dotted black lines in Fig. 4 correspond to the 
    different types of crosses in Fig. 3.
    In either case the previous downward-directed ``nose'' morphs into a (mainly) upward-directed one which is significantly less pronounced, though. Still, there remains one 
    rather dominant peak at 1109. If we were to shift the corresponding minimum to the not less plausible 
    year 1104 (as shown by the bent dotted violet line in Fig. 3), we would end up 
    here at the violet full circle in Fig. 4. 
    Finally, with those corrections in the three intervals we arrive at the rather horizontal 
    black lines in Fig. 4 which are wiggling around a horizontal by not more than 
    $\pm 5$ years.

    In Fig. 5 we show now the curves for Dicke's ratio corresponding to all the
    lines depicted in Fig. 4. This quantity is defined as the ratio $\sum_i^N  r_i^2/\sum_i (r_i-r_{i-1})^2$ 
    between the mean square of the residuals $r_i$ to the mean square of the differences $r_i-r_{i-1}$ between two consecutive residuals (Dicke \citeyear{Dicke1978}). For a random walk process, Dicke's ratio - for $N$ residuals taken into account - behaves as $(N+1)(N^2-1)/(3(5 N^2+6N-3))$ (dark blue line in Fig. 5) with its asymptotic limit $N/15$, while the corresponding dependence for a clocked process reads $(N^2-1)/(2(N^2+2N+3))$ (red line), with its asymptotic limit $1/2$. 
    Note that in Fig. 5 - in contrast to the
    residuals in Fig. 4 - the period is not fixed to 11.07 years, but is separately computed for each value $N$ of data points taken into account. The lowermost green curve shows - again for the Schove/NOAA data - a close proximity to the curve for a clocked process with its asymptotic limit 1/2.
    By contrast, Dicke's ratio  for Usoskin's original data (upper green curve) wiggles around the dark blue curve for a random walk process, in good accordance with the observation by \cite{Weisshaar2023} (who used, though, the slightly different ratio of variances as derived by \cite{Gough1981}). The light blue curve appears after the first correction around 1840, the yellow curve after the second correction around 1650. 
    Hereby, we come already pretty close to
    the theoretical curve for a clocked process (red). The additional corrections between 1040 and 1140 (only shown for the full black curve in Fig. 4) result then in the black curve which is very close to the one for the Schove/NOAA data.

    It is remarkable that the drastic difference between an apparently random-walk like curve such as the light blue one and an apparently clocked-process like curve such as the yellow one stems only from one single additional intervening minimum at 1650.
    If we assume - for the sake of argument - that the additional minimum at 1650 is indeed real, it would just correspond to one single phase jump embedded into an otherwise nicely clocked process, just as discussed recently (referring to other data sets) for two different phase-jump candidates at 1565 and 1795  (Stefani  et al. \citeyear{Stefani2020b}). Showing a strong similarity to a random walk process, the shape of Dicke's ratio would then be completely misleading in this respect. In Fig. A1 of the Appendix we also evidence a strong dependence of the shape of Dicke's ratio on the very position of such an intervening phase jump. It remains to be seen whether an appropriately constructed statistical measure could be found to reliably distinguish between random walk and clocked processes also in case of intervening phase jumps. For the time being we advice to have always a complementary glance on the O-C diagram (such as Fig. 4) which in some sense is an even more telling device than 
    Dicke's ratio.

    Apart from that problem, we also observe that even for the Schove/NOAA  and for the ``optimally corrected'' data of Usoskin Dicke's ratio does not perfectly approach the asymptotic limit 1/2. At this point we reiterate a pertinent argument discussed already in Fig. 2 of \cite{Stefani2019} that a significant share of the variance of the residuals is contained in a long-term cycle of the Suess-de Vries type. This is again illustrated in Fig. 6a which shows the residuals of the Schove/NOAA data - this time for the enlarged interval from 980 to 2009, together with an optimal sinusoidal fit whose period turns out to be 203 years. It is evident that this Suess-de Vries type cycle entails quite a lot of the variance of the data. As a side remark: this outcome speaks much in favour of the quality of Schove's data who had - in the construction of his series - never ``put in''  any long-term periodicity of this kind
    (though he had well recognized it in hindsight, see p. 25 of \cite{Schove1983}).
    When subtracting the fitted 203-yr cycle from the original data, the remaining residuals produce a Dicke ratio as shown as the lower green curve (with triangles) in Fig. 6b, which now much better approaches the asymptotic value of 1/2. 
    
\begin{figure}
   \centering
   \includegraphics[width=0.49\textwidth]{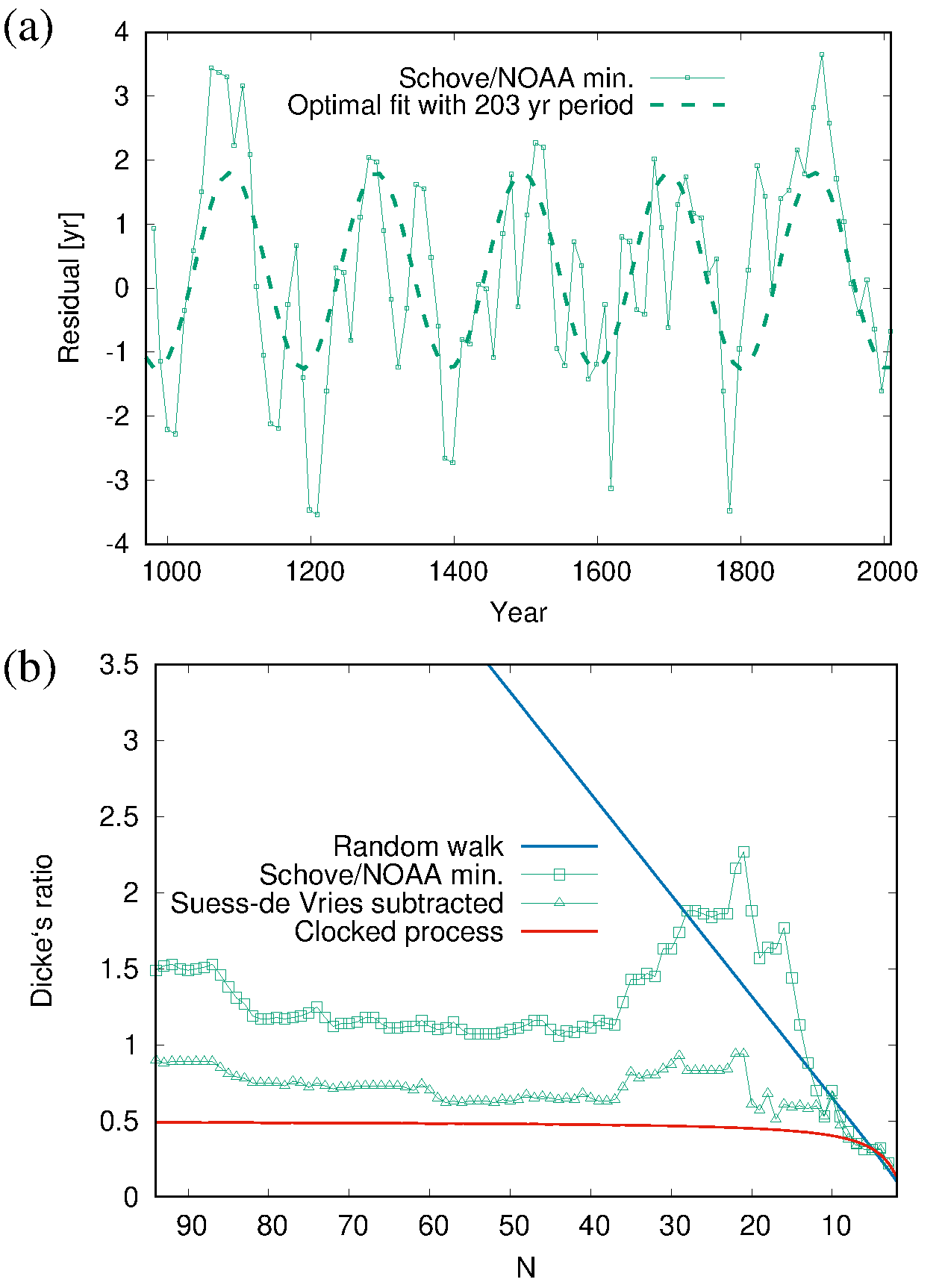}
   \caption{Illustration of the influence of the Suess-de Vries cycle on Dicke's ratio. (a) Residuals for the minima of the combined Schove/NOAA data for the extended interval until 2009. The dashed thick line represents an optimal fit of the residuals with a period of 203 years.
   (b) Dicke's ratio for the residuals of the data in (a), and for the corresponding data with the Suess-de Vries trend being subtracted beforehand. Evidently, after subtraction of the
   Suess-de Vries trend, the approachment of the curve towards the asymptotic limit 0.5 (for clocking) becomes significantly closer. Note in particular that the strong ``overshooting'' of the original Schove/NOAA curve for low $N$ is widely suppressed by this subtraction.}
              \label{FigGam6}%
    \end{figure}

    \section{Conclusions}

    In this paper, we have reanalyzed the series of annual
    (pseudo) sunspot numbers from  \cite{Usoskin2021} with particular view 
    on a possible phase stability of the minima and maxima of the solar cycle.
    
    The corresponding sequence of extrema was compared with another
    sequence comprising Schove's data until 1609 and the standard Schwabe cycles after that year. We have basically confirmed the outcome of \cite{Weisshaar2023} by showing that the curve of Dicke's ratio for the original
    sequence of Usoskin's minima looks formally similar to that of a classical random walk process. Yet, we have also shown that this series
    comprises a very phase stable segment interval between 1140 and 1640, with a one-to-one match of the corresponding extrema with those of the Schove/NOAA series.
    Given that there is only one possibility to get such a one-to-one match,
    in contrast to quite a number of possibilities to infer less or more extrema
    from the original $^{14}$C data, this is already a remarkable result that 
    also reassures the plausibility of Schove's cycle reconstruction (which is
    by some ``regarded as archaic'' \cite{Usoskin2017}).
    
    A first correction of Usoskin's series in form of a contraction of two minima around 1840 into one seems highly plausible given the low quality flag of the former of the two minima and also  in view of the high observational validity of only one minimum in this time span.

    We have tried a second correction in form of a cancellation of one minimum at 1646. Here, amidst the Maunder minimum, the quality flags of all of Usoskin's 
    minima are typically quite low. Admittedly, during this time 
    the observational validity of the standard Schwabe cycle is also not very high, so that the justification of this cancellation remains  doubtful. If we accept it for the moment (also considering that two successive short cycles are not very likely in this particular 
    time of a very {\it weak} solar dynamo) we end up with a
    phase-stable time interval between 1140 and 1890, whose Dicke ratio 
    approaches closely the curve for a clocked process. 
    But even if the additional minimum at 1646 turned 
    out to be real, it would 
    just correspond to one single phase jump embedded into a long period that is otherwise phase stable. This property is best observed in the O-C diagram, while 
    the curve of Dicke's ratio makes the data look like a random walk process.

    The most ambiguous time interval is that between 1040 and 1140
    (basically the Oort minimum), where nearly all quality flags 
    of Usoskin's data are 
    low. The O-C diagram exhibits here a pronounced downward-pointing ``nose'', stemming
    from two very long cycles in the later segment and some correspondingly
    short cycles in the earlier segment of this time interval.
    With two plausible insertions of minima in the later segment, and
    two compensating contractions/cancellations of minima in the earlier segment,
    we reach again a reasonable one-to-one match with the minima of Schove.
    The resulting O-C diagram is significantly smoothed and shows now 
    a pretty horizontal 
    line between 970 and 1890, with
    not much more wiggling (approximately $\pm 5$ years) than in case of the 
    Schove/NOAA data. In our view, this finding strongly reinforces the validity of Schove's data, and impugns Nataf's criticism of them as being simply construed by the ``nine-per-century'' rule.

    Having been focused on a minimal number of corrections pointing (somewhat biasedly) 
    {\it towards} a clocked process, 
    we admit that the high ambiguity of 
    Usoskins's extrema data (with 29 of them having a quality flag $q=0$)
    also entails the possibility that even  {\it more} phase jumps might exist.
    Obviously, with an increasing number of such events the entire notion 
    of phase stability would become more and more problematic.

    At any rate, we conclude that, before entering into a statistical analysis of the 
    clocked - or non-clocked - character of the solar dynamo, the underlying data should be carefully scrutinized. The data set of cycle extrema as produced by 
    \cite{Usoskin2021} entails quite a couple of intervals with low quality flags where the specific extreme should be taken with a grain of salt. While this was clearly expressed in \cite{Usoskin2021}, a too uncritical adoption of the data as by \cite{Weisshaar2023} and 
    \cite{Biswas2023}  might lead to wrong conclusions. In this sense, their argument for a non-clocked process appears premature since its allegedly high significance depends 
    crucially on the selection of the specific set of extrema according to {\cite{Usoskin2021}}.
    While we still refrain from claiming perfect evidence {\it for} solar cycle synchronization, we argue that 
    the work of \cite{Weisshaar2023} does neither represent any conclusive evidence {\it for its absence}.

\begin{acknowledgements}
      This work has received funding from the European Research Council (ERC) under the European Union's Horizon 2020 research and innovation programme (grant agreement No 787544). We thank Carlo Albert and Antonio Ferriz Mas for fruitful discussions about the
      solar dynamo and its possible synchronization.
\end{acknowledgements}

%
%

\begin{appendix} 

\section{Role of the position of a phase jump}

In this appendix we illustrate the influence of an intervening 
additional minimum - embedded into an otherwise clocked process -
on the shape of the curves of Dicke's ratio. For that purpose,
we utilize the Schove/NOAA minima data, this time extended until 2009.

Fig. A.1 shows Dicke's ratio for this data set, and for 6 further data
where phase jumps are artificially inserted at appropriate positions 
close to the years 1000, 1200, 1400, 1600, 1800, and 2000.
Obviously, the similarity of Dicke's ratio to that of a clocked process
becomes most pronounced for phase jumps inserted in the center
of the time interval.

\begin{figure}[h]
\centering
   \includegraphics[width=0.49\textwidth]{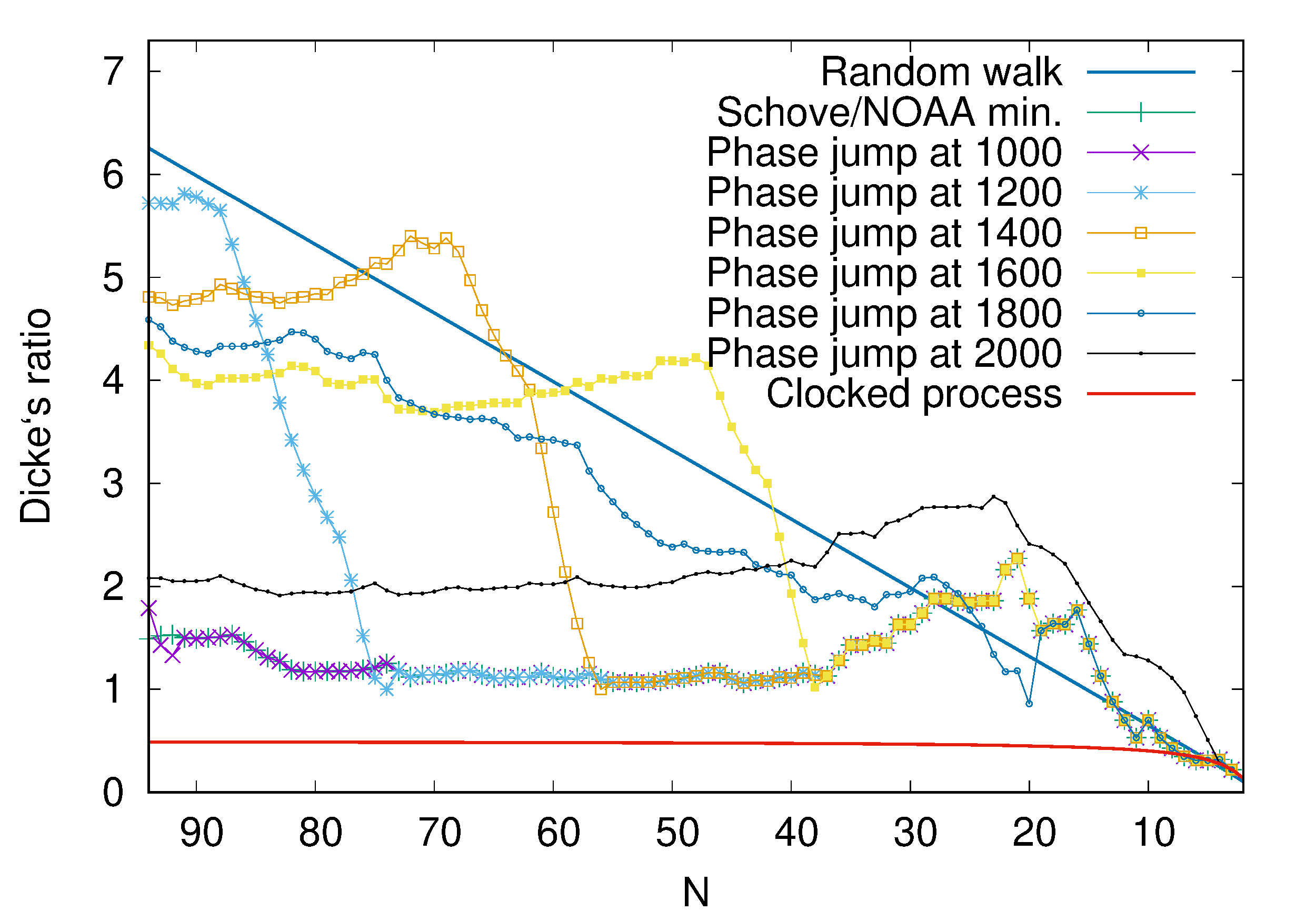}
\caption{Dicke's ratio in presence of the position of phase jumps, based on the Schove/NOAA minimum data between 970 and 2009. Phase jumps are artificially inserted at appropriate positions close to the years
1000, 1200, 1400, 1600, 1800, and 2000.}
\label{FigGam7}
\end{figure}

\end{appendix}

\end{document}